\documentclass[twocolumn,english]{revtex4}
\usepackage[T1]{fontenc}
\usepackage[latin9]{inputenc}
\usepackage{color}
\usepackage{amsbsy}
\usepackage{amstext}
\usepackage{graphicx}

\makeatletter

\newcommand*\LyXZeroWidthSpace{\hspace{0pt}}
\providecommand{\tabularnewline}{\\}

\@ifundefined{textcolor}{}
{%
 \definecolor{BLACK}{gray}{0}
 \definecolor{WHITE}{gray}{1}
 \definecolor{RED}{rgb}{1,0,0}
 \definecolor{GREEN}{rgb}{0,1,0}
 \definecolor{BLUE}{rgb}{0,0,1}
 \definecolor{CYAN}{cmyk}{1,0,0,0}
 \definecolor{MAGENTA}{cmyk}{0,1,0,0}
 \definecolor{YELLOW}{cmyk}{0,0,1,0}
}

\makeatother

\usepackage{babel}
\begin{document}
\title{An efficient method for strongly correlated electrons in two-dimensions}
\author{Ion Mitxelena$^{1}$ and Mario Piris$^{1,2}$\bigskip{}
}
\address{$^{1}$Kimika Fakultatea, Euskal Herriko Unibertsitatea (UPV/EHU)
and Donostia International Physics Center (DIPC), 20018 Donostia,
Euskadi, Spain. \\
$^{2}$Basque Foundation for Science (IKERBASQUE), 48013 Bilbao, Euskadi,
Spain.\bigskip{}
}
\begin{abstract}
This work deals with the problem of strongly correlated electrons
in two-dimensions (2D). We give a reduced density matrix (RDM) based
tool through which the ground-state energy is given as a functional
of the natural orbitals and their occupation numbers. Specifically,
the Piris Natural Orbital Functional 7 (PNOF7) is used for studying
the 2D Hubbard model and hydrogen square lattices. The singlet ground-state
is studied, as well as the doublet mixed quantum state obtained by
extracting an electron from the system. Our method satisfies two-index
necessary N-representability conditions of the two-particle RDM (2RDM)
and guarantees the conservation of the total spin. We show the ability
of PNOF7 to describe strong correlation effects in 2D systems
by comparing our results with exact diagonalization, density matrix
renormalization group (DMRG), and auxiliary-field quantum Monte Carlo
calculations. PNOF7 overcomes variational 2RDM methods with two- and
three-index positivity N-representability conditions, reducing computational
cost to mean-field scaling. Consistent results are obtained for small
and large systems up to 144 electrons, weak and strong correlation
regimes, and many filling situations. Unlike other methods, there
is no dependence on dimensionality in the results obtained with PNOF7,
and no particular difficulties have been observed to converge PNOF7
away from half-filling. Smooth double occupancy of sites is obtained
regardless of the filling. Symmetric dissociation of 2D hydrogen lattices
shows that long-range nondynamic correlation drammatically affects
electron detachment energies. PNOF7 compares well with DMRG along
the dissociation curve.\bigskip{}

\end{abstract}
\maketitle

\section{Introduction}

The Hubbard model defined on two-dimensional (2D) lattices with repulsive
interactions constitutes a powerful tool to understand the physics
of 2D materials. For instance, the 2D Hubbard model has been recently
employed to describe experimental observations in graphene nanoribbons
\cite{materials-2}, as well as to study the phase diagram of high-$T_{c}$
cuprate superconductors \cite{Zheng1155}. A general solution for
the 2D Hubbard model remains unknown, although several approaches
have proven to be accurate for specific cases, namely the auxiliary-field
quantum Monte Carlo (AFQMC) method at half-filling \cite{Mingpu},
the density matrix embedding theory (DMET) \cite{dmet-prb} in the
noninteracting and atomic limits, and the density matrix renormalization
group (DMRG) algorithm \cite{renormalization-group} if a sufficient
number of retained renormalized basis states is considered \cite{Verstichel2014}.
The latter is probably the most efficient method to study one-dimensional
(1D) systems, however, the DMRG performance is less accurate in 2D
due to its 1D topology \cite{dmrg1-PRB}.

A recent benchmarking \cite{PHYSREVX_2Dhubbard} shows the performance
of well-established methods in quantum chemistry in the context of
2D fermionic systems. Many of these methods, such as the coupled cluster
singles and doubles with perturbative triples (CCSD(T)), the gold-standard
in quantum chemistry, dramatically fail at strong correlation regimes.
Variational methods based on reduced density matrices
(RDM) emerge as promising alternatives for studying strongly correlated
materials \cite{Pernal_2015_new_journal}. The variational second-order
RDM (v2RDM) method has demonstrated to accurately describe the 2D
Hubbard model at half-filling \cite{ANDERSON201322} and away from
half-filling \cite{Verstichel2014} if three-index constraints are
imposed. Unfortunately, this implies a computational scaling
of $\mathcal{O}\left(M^{9}\right)$, $M$ being the dimension of the
single-particle space.

Favorable computational efficiency can be achieved using one-particle
theories such as density functional theory (DFT) or first-order RDM
(1RDM) functional theory (1RDMFT) \cite{Pernal2016}. Despite recent
efforts made \cite{becke_dft_strong,Su_DFT_strong} to describe the
strong electron correlation, DFT in its conventional local or semilocal
approximations is still unable to provide a correct description of
correlated insulators \cite{Carrascal}. Conversely, 1RDMFT describes
correctly metal insulator transitions \cite{Shinohara_2015} and strongly
correlated electrons in 1D \cite{Mitxelena2017a,Mitxelena2018b,mitxelena2018a,Schilling2019,MITXELENA2019}.
The goal of this paper is to investigate the ability of 1RDMFT to
deal with correlated electrons in 2D systems. Recently \cite{Saubanere-prb-2011},
a formulation of 1RDMFT defined on a lattice has been published, based
on the exact solution of the two-site problem. Along this paper we demonstrate that
a more general and fundamental formulation of 1RDMFT is equally valid
to deal with lattice models, without any loss of generality.

In most applications, the spectral decomposition of the 1RDM is used
to express it in terms of the naturals orbitals (NOs) and their occupation
numbers (ONs). In this representation, the energy expression is referred
to as natural orbital functional (NOF) \cite{Piris2007,Piris2014a}.
In this paper, we provide an extensive study of 2D systems using the
Piris NOF 7 (PNOF7) \cite{Piris2017,mitxelena2018a} and
the formulation for spin-multiplets \cite{Piris2019}, which are reviewed
in section \ref{sec:Natural-Orbital-Functional}. In section \ref{sec:Model-systems},
we employ the 2D Hubbard model varying the relative contribution between
the hopping ($t$) and electron-electron on-site interaction ($U$)
parameters, as well as the filling, for many system sizes that
reach up to 12x12 square lattices. Here, our results are compared
to state-of-the-art methods for the study of strong correlation, such
as DMRG, v2RDM, AFQMC, and exact diagonalization (ED).

\textcolor{black}{The lack of long-range inter-electronic interactions
may be the most important limitation of the Hubbard model. In fact,}
it has been recently \cite{review_Hchain} emphasized that the properties
of hydrogen chains can strongly differ from those obtained by means
of the 1D Hubbard model. Accordingly, in section \ref{sec:Two-dimensional-hydrogen-lattice}
we focus on 2D lattices of hydrogen atoms to \textcolor{black}{model
the strong electron correlation in the presence of long-range interaction
effects. }The symmetric dissociation of 2D hydrogen lattices is studied.

\section{\label{sec:Natural-Orbital-Functional}Natural Orbital Functional
for Multiplets}

In order to obtain an approximate NOF, the electronic energy is usually
given in terms of the NOs $\left\{\phi_{i}\right\}$ and their ONs $\left\{n_{i}\right\}$ as follows
\begin{equation}
E=\sum\limits _{i}n_{i}\mathcal{H}_{ii}+\sum\limits _{ijkl}D[n_{i},n_{j},n_{k},n_{l}]<kl|ij>\label{NOF}
\end{equation}
where $\mathcal{H}_{ii}$ denotes the diagonal elements of the one-particle
part of the Hamiltonian involving the kinetic energy and the external
potential operators, $<kl|ij>$ are the matrix elements of the two-particle
interaction, and $D[n_{i},n_{j},n_{k},n_{l}]$ represents the reconstructed
2RDM from the ONs. Restriction of the ONs to the range $0\leq n_{i}\leq1$
represents a necessary and sufficient condition for ensemble $\mathrm{N}$-representability
of the 1RDM \cite{Coleman1963} under the normalization condition
$\sum_{i}n_{i}=\mathrm{N}$. Recall that there is no approximation
in the functional dependencies involving one-electron operators, so
the latter are exactly described.

Let us consider a non-relativistic Hamiltonian free of
spin coordinates, hence the ground state with total spin $S$ is a
multiplet, i.e., a mixed quantum state (ensemble) that allows all
possible $S_{z}$ values. Next, we briefly describe how we do the
reconstruction of $D$ to achieve the PNOF7 for spin-multiplets. A
more detailed description can be found in Ref. \cite{Piris2019}. 

First, we consider $\mathrm{N_{I}}$ single electrons and $\mathrm{N_{II}}$
paired electrons, so that $\mathrm{N_{I}}+\mathrm{N_{II}}=\mathrm{N}$.
We also assume that all spins corresponding to $\mathrm{N_{II}}$
electrons are coupled as a singlet, thence the $\mathrm{N_{I}}$ electrons
determine the spin $S$ of the system. We focus on the mixed state
of highest multiplicity: $2S+1=\mathrm{N_{I}}+1,\,S=\mathrm{N_{I}}/2$,
$\mathrm{<}\hat{S}^{2}\mathrm{>}=\mathrm{N_{I}}/2\left(\mathrm{N_{I}}/2+1\right)$.
For this ensemble of pure states $\left\{ \left|SM\right\rangle \right\} $,
the expected value of $\hat{S}_{z}$ is zero, namely, 
\begin{equation}
\mathrm{<}\hat{S}_{z}\mathrm{>}=\frac{1}{\mathrm{N_{I}}+1}{\textstyle {\displaystyle \sum_{M=-\mathrm{N_{I}}/2}^{\mathrm{N_{I}}/2}}M}=0
\end{equation}

Consequently, we can adopt the spin-restricted theory in which a single
set of orbitals $\left\{\varphi_{p}\right\}$ is used for $\alpha$ and $\beta$ spin projections. All spatial
orbitals $\varphi_{p}$ will be then doubly occupied in the ensemble, so that occupancies
for particles with $\alpha$ and $\beta$ spins are equal: $n_{p}^{\alpha}=n_{p}^{\beta}=n_{p}.$

In turn, we divide the orbital space $\Omega$ into two subspaces:
$\Omega=\Omega_{\mathrm{I}}\oplus\Omega_{\mathrm{II}}$. The orbital
space $\Omega_{\mathrm{II}}$ is composed of $\mathrm{N_{II}}/2$
mutually disjoint subspaces: $\Omega_{\mathrm{II}}=\Omega_{1}\oplus\Omega_{2}\oplus...\oplus\Omega_{\mathrm{N_{II}}/2}$,
$\Omega_{f}\cap\Omega_{g}=\textrm{Ø}$ \citep{Piris2018a}. Note that the subscripts $f$ and $g$ are used from now on to refer 
to different subspaces. 

Each subspace $\Omega{}_{g}\in\Omega_{\mathrm{II}}$ contains one orbital $g$ below the level $\mathrm{N_{II}}/2$, 
and $\mathrm{N}_{g}$ orbitals above it. In general, $\mathrm{N}_{g}$ may be different for each $\Omega{}_{g}$ subspace, but for simplicity 
we take all equal. The maximum value allowed for $\mathrm{N}_{g}$ is determined by the basis set used in the calculation. 
In this paper, we restrict to the minimal basis approach, so $\mathrm{N}_{g}$ is equal to 1 in all calculations.

Taking into account the spin, the total occupancy for a given subspace $\Omega{}_{g}\in\Omega{}_{\mathrm{II}}$ is 2, which is reflected
in additional sum rule, namely,
\begin{equation}
\sum_{p\in\Omega_{g}}n_{p}=1,\,\Omega{}_{g}\in\Omega{}_{\mathrm{II}}\label{sum1}
\end{equation}
It follows that
\begin{equation}
2\sum_{p\in\Omega_{\mathrm{II}}}n_{p}=2\sum_{g=1}^{\mathrm{N_{II}}/2}\sum_{p\in\Omega_{g}}n_{p}=\mathrm{N_{II}}\label{sumNp}
\end{equation}

Similarly, $\Omega_{\mathrm{I}}$ is composed of $\mathrm{N_{I}}$
mutually disjoint subspaces $\Omega{}_{g}$, but in contrast to $\Omega_{\mathrm{II}}$,
each subspace $\Omega{}_{g}\in\Omega_{\mathrm{I}}$ contains only
one orbital $g$ with $2n_{g}=1$. It is worth noting that each orbital
is completely occupied individually, but we do not know whether the
electron has $\alpha$ or $\beta$ spin: $n_{g}^{\alpha}=n_{g}^{\beta}=n_{g}=1/2$. Accordingly
\begin{equation}
2\sum_{p\in\Omega_{\mathrm{I}}}n_{p}=2\sum_{g=\mathrm{N_{II}}/2+1}^{\mathrm{N_{II}}/2+\mathrm{N_{I}}}n_{g}=\mathrm{N_{I}}\label{sumNpp}
\end{equation}
Note that $p=g$ if $p\in\Omega_{I}$ since there is only one orbital in that subspace.
Taking into account eqs. (\ref{sumNp}) and (\ref{sumNpp}), the trace of the 1RDM is
verified equal to the number of electrons
\begin{equation}
2\sum_{p\in\Omega}n_{p}=2\sum_{p\in\Omega_{\mathrm{II}}}n_{p}+2\sum_{p\in\Omega_{\mathrm{I}}}n_{p}=\mathrm{N_{II}}+\mathrm{N_{I}}=\mathrm{\mathrm{N}}
\end{equation}
To guarantee the existence of an N-electron system compatible with
the functional (\ref{NOF}), we must observe the N-representability
conditions \cite{Mazziotti2012a} on the reconstructed 2RDM \cite{Piris2018}.
Assuming real spatial orbitals, the employment of necessary N-representability
conditions leads to PNOF7 for multiplets
\begin{equation}
E=\sum\limits _{g=1}^{\mathrm{N_{II}}/2}E_{g}+\sum\limits _{g=\mathrm{N_{II}}/2+\mathrm{1}}^{\mathrm{N_{II}}/2+\mathrm{N_{I}}}\mathcal{H}_{gg}+\sum\limits _{f\neq g=1}^{\mathrm{N_{II}}/2+\mathrm{N_{I}}}E_{fg}\label{EPNOF7}
\end{equation}
\begin{equation}
E_{g}=2\sum\limits _{p\in\Omega_{g}}n_{p}\mathcal{H}_{pp}+\sum\limits _{q,p\in\Omega_{g}}\Pi_{qp}\mathcal{K}_{pq},\,\Omega{}_{g}\in\Omega{}_{\mathrm{II}}
\end{equation}
\begin{equation}
\Pi_{qp}=\left\{ \begin{array}{c}
\sqrt{n_{q}n_{p}}\,,\quad q=p\textrm{ or }q,p>\frac{N_{\mathrm{II}}}{2}\\
-\sqrt{n_{q}n_{p}}\,,\quad q=g\textrm{ or }p=g\qquad\;
\end{array}\right.
\end{equation}
\begin{equation}
E_{fg}=\sum\limits _{p\in\Omega_{f}}\sum\limits _{q\in\Omega_{g}}\left[n_{q}n_{p}\left(2\mathcal{J}_{pq}-\mathcal{K}_{pq}\right)-\Phi_{q}\Phi_{p}\mathcal{K}_{pq}\right]
\end{equation}
where $\Phi_{p}=\sqrt{n_{p}(1-n_{p})}$. $\mathcal{J}_{pq}$ and $\mathcal{K}_{pq}$
refer to the usual Coulomb and exchange integrals $\left\langle pq|pq\right\rangle $
and $\left\langle pq|qp\right\rangle $, respectively. It should be
noted that $E_{g}$ reduces to a NOF obtained from ground-state singlet
wavefunction, so it describes accurately two-electron systems \cite{Piris2018a}.
In the last term of eq. (\ref{EPNOF7}), $E_{fg}$ correlates the
motion of electrons with parallel and opposite spins belonging to
different subspaces ($\Omega_{f}\neq\Omega{}_{g}$).

\textcolor{black}{The solution is established by optimizing the energy
(}\ref{EPNOF7}\textcolor{black}{) with respect to the ONs and to
the NOs, separately. The conjugate gradient method is used for performing
the optimization of the energy with respect to auxiliary variables
that enforce automatically the N-representability bounds of the 1RDM.
The self-consistent procedure proposed in \cite{Piris2009a} yields
the NOs by an iterative diagonalization procedure, in which }orbitals
are not constrained to remain fixed along the orbital optimization
process. All calculations have been carried out by using the DoNOF
code developed in our group.

\section{\label{sec:Model-systems}Two-Dimensional Hubbard model}

The Hubbard model is the simplest prototype for modeling strongly
correlated systems. Its Hamiltonian reads as

\begin{equation}
\hat{H}=-t\sum_{\langle\boldsymbol{r},\boldsymbol{r'}\rangle,\sigma}(\hat{a}_{\boldsymbol{r},\sigma}^{\dagger}\hat{a}_{\boldsymbol{r'},\sigma}+\hat{a}_{\boldsymbol{r'},\sigma}^{\dagger}\hat{a}_{\boldsymbol{r},\sigma})+U\sum_{\boldsymbol{r}}\hat{n}_{\boldsymbol{r},\alpha}\hat{n}_{\boldsymbol{r},\beta}\label{HH}
\end{equation}
where $\hat{a}_{\boldsymbol{r},\sigma}^{\dagger}\left(\hat{a}_{\boldsymbol{r},\sigma}\right)$
is the creation (annihilation) operator, so the first term allows
hopping between nearest neighboring sites $\boldsymbol{r}$ and $\boldsymbol{r'}$.
$\sigma=\alpha,\beta$ stands for the spin, and $\hat{n}_{\boldsymbol{r},\sigma}=\hat{a}_{\boldsymbol{r},\sigma}^{\dagger}\hat{a}_{\boldsymbol{r},\sigma}$
gives the number of electrons on site $\boldsymbol{r}$ with spin
$\sigma$. Let us restrict to the 2D model, so each vector \textbf{r}
has two components. In the noninteracting limit ($U=0$), the Hartree-Fock
2RDM provides the exact solution for the tight-binding Hamiltonian,
while for nonzero electron-electron on-site interactions ($U\neq0$)
a correlated approximation for the 2RDM must be given. Hereafter,
$U/t$ will be employed to quantify the bridge from the noninteracting
limit ($U/t\rightarrow0$) to the atomic or strong correlation limit
($U/t\rightarrow\infty$), also referred to as metal to insulator
transition.

In the following, we test the performance of PNOF7 against some of
the benchmarks used in Refs. \cite{PHYSREVX_2Dhubbard,Zheng1155}.
The results obtained by means of v2RDM with\textcolor{red}{{}
}two-index constraints (PQG), and $T1$ and $T2'$ N-representability
conditions (PQGT') are also included. We study the Hubbard model on
2D square lattices for different sizes, filling situations (or densities),
and spin multiplicities. Periodic boundary conditions are employed in all directions.

\subsection{Half-filling}

Let us set the number of electrons to be equal to the number of sites,
so we have a half-filled lattice. There is one electron per
site, that is, half the maximum possible number (two electrons per site). 

\begin{figure}[h]
\includegraphics[scale=0.99]{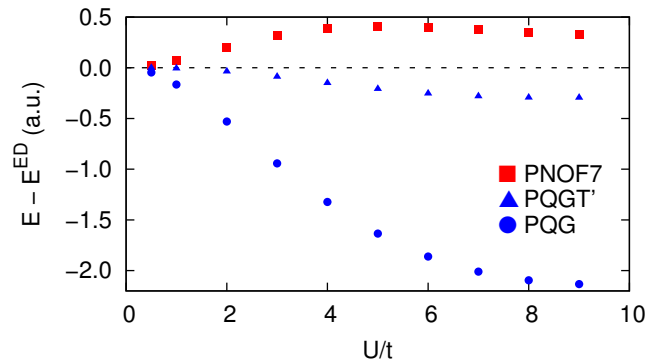}

\caption{\label{figures-4x4-2Dhubbard}PNOF7, PQG, and PGQT' energy differences at
half-filling for the 4x4 square lattice Hubbard model. ED, PQG, and
PQGT' data from Ref. \cite{ANDERSON201322}.}
\end{figure}

In Fig. \ref{figures-4x4-2Dhubbard}, we show the PNOF7, PQG, and PGQT' energy differences
with respect to the ED values for the 4x4 square lattice Hubbard
model at half-filling. Fig. \ref{figures-4x4-2Dhubbard}
reveals that PNOF7 is in good agreement with the ED results and perform
similar to PQGT'. The latter reproduces accurately the ED results, but
at the expense of an unfavorable scaling of $\mathcal{O}\left(M^{9}\right)$.
The scaling can be reduced to $\mathcal{O}\left(M^{6}\right)$ if
only two-index constraints are applied, but this deteriorates the
performance for large $U/t$ as shown by the PQG results. 

\begin{figure}[h]
\includegraphics[scale=0.99]{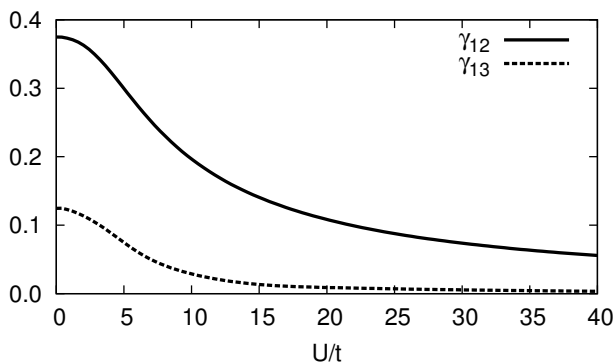}

\caption{\label{figures-4x4-2Dhubbard-gamma}PNOF7 1RDM in the site basis at
half-filling for the 4x4 square lattice Hubbard model.}
\end{figure}

It is interesting to look at the 1RDM in the site basis, denoted hereafter as $\gamma$. 
Since PNOF7 contains particle-hole symmetry, $\gamma$ is completely symmetric 
(i.e. $\gamma_{ij}=\gamma_{ji}$). All sites are equivalent, so we have just to look at the 
elements involving one site and its neighbours. At half-filling, the average occupation 
of each site is one for any value of $U/t$, i.e. $\gamma_{11}=1.0$. Conversely, off-diagonal 
elements vary depending on the correlation regime. As shown in Fig. \ref{figures-4x4-2Dhubbard-gamma}, 
the largest value is obtained for nearest-neighbours, so that $\gamma_{12}$ is maximum at $U/t=0$, 
and it decreases monotonically to zero at the strong correlation limit. The latter is intimately 
related with electron delocalization, which is inversely proportional to $U/t$. Fig. 
\ref{figures-4x4-2Dhubbard-gamma} also includes $\gamma_{13}$, which is another 
non-vanishing off-diagonal element that shows a similar dependence on electron correlation. 
As expected, these values are four times degenerated due to the 2D nature of the system 
with periodic boundary conditions in both directions.

\begin{table}[h]
\centering{}\caption{\label{table_2D_hubbard_energies}PNOF7 and exact energies (in a.u.) from AFQMC
calculations \cite{Mingpu} for the Hubbard model defined on 8x8, 10x10, and 12x12
square lattices at half-filling varying $U/t$. Mean absolute
error per site (MAES) is shown for the three systems.\smallskip{}
}
\begin{tabular}{|c||c|c|c|c|c|c|}
\hline 
$U/t$ & \multicolumn{2}{c|}{8x8} & \multicolumn{2}{c|}{10x10} & \multicolumn{2}{c|}{12x12}\tabularnewline
\hline 
\hline 
 & Exact & PNOF7 & Exact & PNOF7 & Exact & PNOF7\tabularnewline
\hline 
2 & -74.47 & -73.37 & -116.91 & -115.19 & -168.75 & -166.17\tabularnewline
\hline 
4 & -55.05 & -53.27 & -86.12 & -83.32 & -123.95 & -120.20\tabularnewline
\hline 
6 & -42.16 & -40.53 & -64.80 & -63.39 & -94.66 & -91.24\tabularnewline
\hline 
8 & -33.68 & -32.26 & -52.54 & -50.48 & -75.54 & -72.53\tabularnewline
\hline 
\hline 
MAES & \multicolumn{2}{c|}{0.02} & \multicolumn{2}{c|}{0.02} & \multicolumn{2}{c|}{0.02}\tabularnewline
\hline 
\end{tabular}
\end{table}

At half-filling, the AFQMC method turns out to be numerically exact \cite{Mingpu},
so it can be used as benchmark in larger systems. In Table \ref{table_2D_hubbard_energies},
we show the absolute energies corresponding to the 8x8 = 64, 10x10 = 100,
and 12x12 = 144 sites 2D Hubbard model. It should be
noted that the conclusions obtained in Ref. \cite{mitxelena2018a}
for the 1D Hubbard model hold in two-dimensions: (1) results obtained
by using PNOF7 are comparable to exact results for a wide range of
$U/t$ values, and (2) if we consider the mean absolute
error per site (MAES), the performance of PNOF7 does not deteriorate
with the increasing size of the system. According to Table \ref{table_2D_hubbard_energies},
PNOF7 recovers approximately 98\% of the total energy for $U/t=2$,
whereas for $U/t=8$, 96\% of the total energy is retrieved. The MAES does not change 
significantly going from 64 to 144 sites, in fact, it is approximately equal to 0.02 for all systems studied.
Note that the strong correlation limit at half-filling has not been reached for the $U/t$ values 
reported in Table I (vide infra).

\begin{table}[h]
\centering{}\caption{\label{EDE_10x10}PNOF7 electron detachment energies (EDE), in a.u.,
for the 10x10 square lattice Hubbard model at half-filling.\medskip{}
}
\begin{tabular}{|c||c|c|c|c|}
\hline 
$U/t$ & 2 & 4 & 6 & 8\tabularnewline
\hline 
EDE & -0.48 & -1.06 & -1.61 & -1.88\tabularnewline
\hline 
\end{tabular}
\end{table}

In Table \ref{EDE_10x10}, we report the PNOF7 electron detachment
energies (EDE) for the 10x10 square lattice by varying $U/t$. The EDEs
are computed as the energy difference between doublet- and singlet-spin
states ($E^{S=1/2}-E^{S=0}$), so that an electron is removed to produce
$S=1/2$. We observe that EDEs are negative since in our model Hamiltonian
(\ref{HH}) we do not consider an on-site attractive potential that
represents the effects of an external field on the electrons.
EDE increase in absolute value as the on-site electron-electron repulsion
increases for a given $t$, since overall electron repulsion is reduced
by removing an electron. Accordingly, the localization of electrons
that takes place at half-filling with large $U/t$ values favors the
removal of an electron.

\subsection{Away from half-filling}

In Eq. (\ref{EPNOF7}), the last term of PNOF7 correlates the motion
of electrons belonging to different subspaces by introducing explicitly
the particle-hole symmetry through $\Phi_{p}=\sqrt{n_{p}(1-n_{p})}$.
Consequently, PNOF7 is expected to be particularly accurate at half-filling
since the Hubbard model exhibits particle-hole symmetry in this case.
Let us now break the particle-hole symmetry of the system, so that
inhomogeneous phases can appear in the ground state \cite{PHYSREVX_2Dhubbard},
in order to test the performance of PNOF7 away from the half-filling.
Breaking this symmetry strongly affects the nature of the system,
since it deformates its Fermi surface and, therefore, the corresponding electronic
interactions turn out to be less localized \cite{dmrg1-PRB}.

\begin{figure}[h]
\includegraphics[scale=0.99]{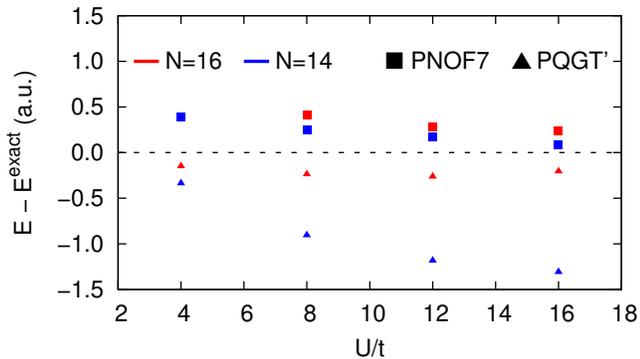}

\caption{\label{figures-4x4-2Dhubbard-1}PNOF7 and PQGT' energies with respect
to ED for the 4x4 square lattice Hubbard model with 14 and 16 electrons.
ED and PQGT' data from \cite{Verstichel2014}.}
\end{figure}
\begin{figure}[h]
\includegraphics[scale=0.99]{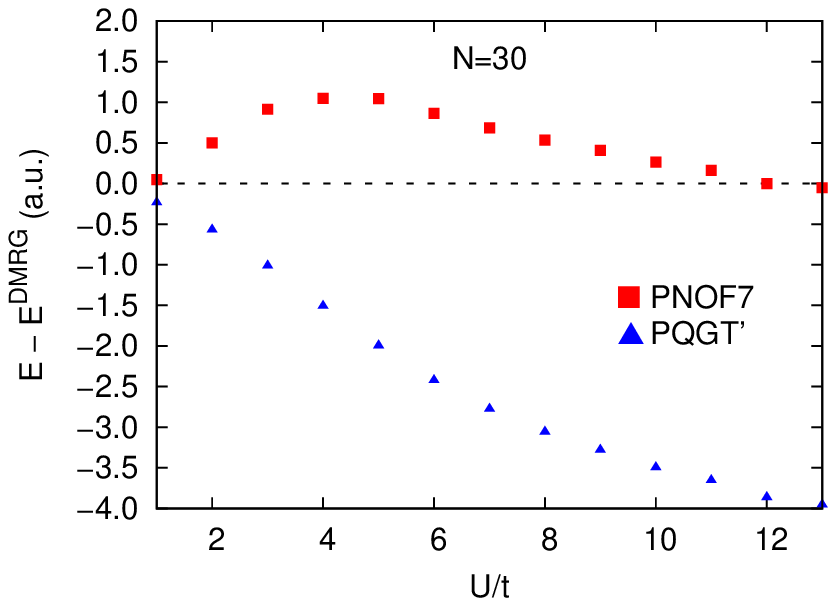}

\includegraphics[scale=0.99]{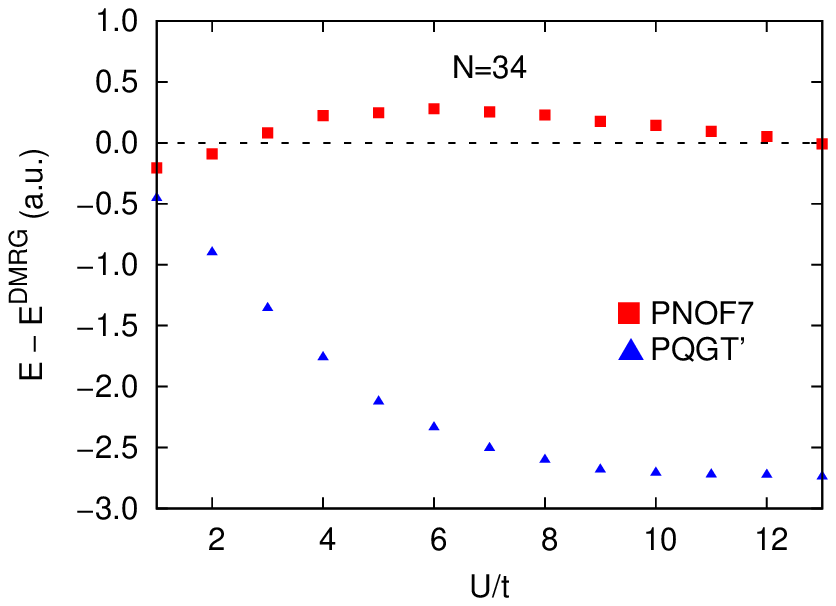}

\includegraphics[scale=0.99]{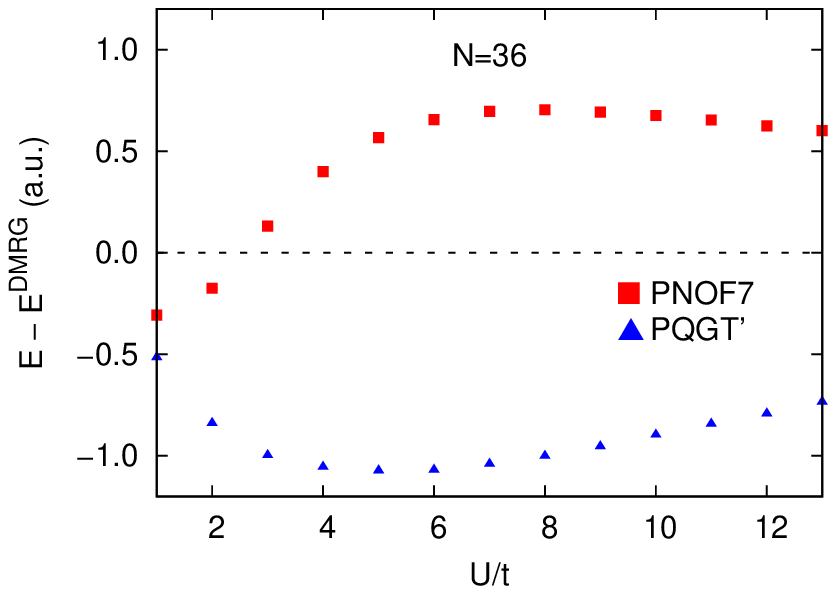}

\caption{\label{figures-6x6-2Dhubbard}PNOF7 and PQGT' energy differences
with respect to DMRG for the
6x6 square lattice Hubbard model with 30, 34, and 36 electrons. DMRG
and PQGT' data from \cite{Verstichel2014}.}
\end{figure}

Results corresponding to the 4x4=16 sites Hubbard model including
14 and 16 electrons are shown in Fig. \ref{figures-4x4-2Dhubbard-1}.
Exact results corresponding to ED, as well as approximate energies
from v2RDM with PQGT' constraints, are taken from Ref. \cite{Verstichel2014}. PNOF7 retains
its accuracy for any filling as $U/t$ increases, while v2RDM fails
away from half-filling in the intermediate and strong correlation
regimes. The maximum error shown by PNOF7 ($\Delta E\sim2.5\%$) corresponds
to $U/t=4$, wherein the two points, N=14 and N=16, are on top of
each other. Note that PNOF7 approaches the exact result for large
$U/t$ values.

In Fig. \ref{figures-6x6-2Dhubbard}, we show the PNOF7 and PQGT' energy
differences with respect to DMRG for the 6x6 square lattice Hubbard model at
different correlation regimes and fillings, including N = 30, 34,
and 36 electrons. DMRG and PQGT' values are taken from Ref. \cite{Verstichel2014}.
Although the largest deviations with respect to DMRG are obtained
for the lowest density, corresponding to 30 electrons in 36 sites,
the PNOF7 agreement with DMRG is below 1 a.u. for all $U/t$ values
reported.

At low densities (N=34 or N=30) PNOF7 reaches the strong correlation
limit at smaller $U/t$ values, as expected. In contrast, v2RDM cannot recover
the large amount of correlation energy at partial-filling, although
it does correctly in 1D \cite{verstichel-prl,rubin_mazziotti_2015,Rubin_mazziotti_2014}.
Such dependence on dimensionality does not appear with PNOF7. We must
recall that the N-representability conditions are imposed in the construction
of the functional \cite{Piris2017,mitxelena2018a}, whereas in the v2RDM
methods these constraints are imposed in the minimization of the energy
functional by means of semidefinite programming techniques. The latter
may become numerically unstable if all the states are nearly degenerate
\cite{ANDERSON201322}, something that is observed in strongly correlated
systems. The advantages of imposing N-representability constraints
on the construction of the functional rather than the minimization
process have already been emphasized \cite{Gritsenko_pra2019} for
pure N-representability conditions of the 1RDM. In addition, we have
not observed particular difficulties to converge PNOF7 away from half-filling,
in contrast to many other numerical methods in which convergence errors
arise due to the lack of particle-hole symmetry \cite{PHYSREVX_2Dhubbard}.

\begin{figure}
\includegraphics[scale=0.99]{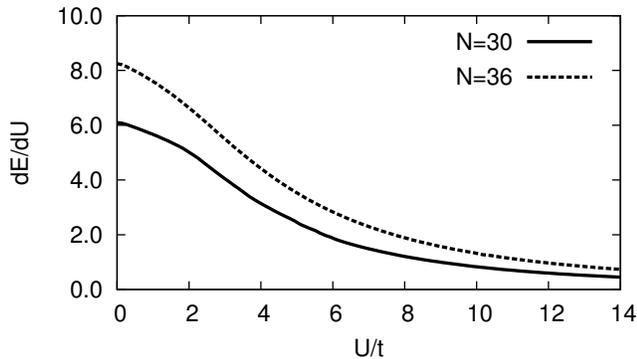}

\caption{\label{double-occupation}Double occupancy as a function of $U/t$
for the 6x6 square lattice Hubbard model with 30 and 36 electrons.}
\end{figure}

To obtain a more reliable indication of the energy, its derivative
with respect to the parameter $U$ was also studied, namely,
\begin{equation}
\frac{dE}{dU}=\sum_{\boldsymbol{r}}\left\langle \hat{n}_{\boldsymbol{r},\alpha}\hat{n}_{\boldsymbol{r},\beta}\right\rangle \label{doubleocc}
\end{equation}

According to Eq. (\ref{doubleocc}), $dE/dU$ yields the double occupancy
of the sites. This magnitude is very sensitive to the NOF used in
the Hubbard model as our previous study on 1D systems demonstrated
\cite{Mitxelena2017a}. Several functionals other than PNOF7 produced
discontinuous curves for double occupancy of the sites. In this paper,
the double occupancy is numerically evaluated by using the formula
\begin{equation}
\frac{dE}{dU}\approx\frac{E(U-2h)-8E(U-h)+8E(U+h)-E(U+2h)}{12h}\label{dEdUapprox}
\end{equation}
where the step size $h$ is set to $10^{-3}$.

In Fig. \ref{double-occupation}, we report the double occupancy of
sites as a function of $U/t$ for the 6x6 square lattice Hubbard model
with 30 and 36 electrons. As expected, the double occupancy is maximum
in the weak correlation region, since there are no two-particle interactions.
The population of the sites spreads out as the correlation increases,
so for large $U/t$ values the double occupancy decreases due to the
electron-electron on-site interaction. As we can see in Fig. \ref{double-occupation},
PNOF7 produces smooth double occupancy without discontinuities for
the 2D Hubbard model. Our method shows qualitatively good trend for
increasing $U/t$ regardless of the filling.

\begin{table}
\centering{}\caption{\label{EDE_6x6}PNOF7 electron detachment energies, in a.u., for the
6x6 square lattice Hubbard model with 30 and 36 electrons.\medskip{}
}
\begin{tabular}{|c||c|c|c|c|}
\hline 
$U/t$ & 1 & 6 & 10 & 14\tabularnewline
%\hline 
\hline 
N=36 & -0.35 & -1.70 & -2.27 & -2.62\tabularnewline
\hline 
N=30 & 0.46 & -1.01 & -1.42 & -1.61\tabularnewline
\hline 
\end{tabular}
\end{table}

In Table \ref{EDE_6x6}, we report PNOF7 EDE for the 6x6 square lattice
with 30 and 36 electrons by varying $U/t$. Table \ref{EDE_6x6} reveals
that the EDE can take positive and negative values at low densities,
in contrast to the half-filling ($N=36$). In fact, at large correlation
regimes EDE are negative, whereas they become positive in presence
of weak correlation effects and low filling, as is the case for $U/t\sim1$
and $N/N_{s}=30/36=0.833$ (where $N_{s}$ represents the number of
sites). Note from Table \ref{EDE_6x6} that the EDE increase in absolute
value together with the amount of correlation, as it has been observed
for the 10x10 lattice at half-filling (see Table \ref{EDE_10x10}).

\section{\label{sec:Two-dimensional-hydrogen-lattice}Two-dimensional hydrogen
lattice}

A hydrogen chain resembles the 1D Hubbard model with long-range interactions
if we employ a minimal basis set \cite{review_Hchain}. These interactions
may produce significant differences between both model systems. Recently
\cite{mitxelena2019c}, we have proven the ability of PNOF7 to describe
the symmetric and asymmetric dissociations of a 1D hydrogen chain
with 50 atoms, wherein the PNOF7 energies compared remarkably well with
those obtained at the DMRG level of theory along the dissociation
curves.

The addition of a spatial dimension increases the amount of interactions
and makes the bond-breaking process more complex, so new and diverse
strong correlations can emerge. This section is dedicated to the study
of 2D hydrogen square lattices.

\begin{figure}[h]
\includegraphics[scale=0.99]{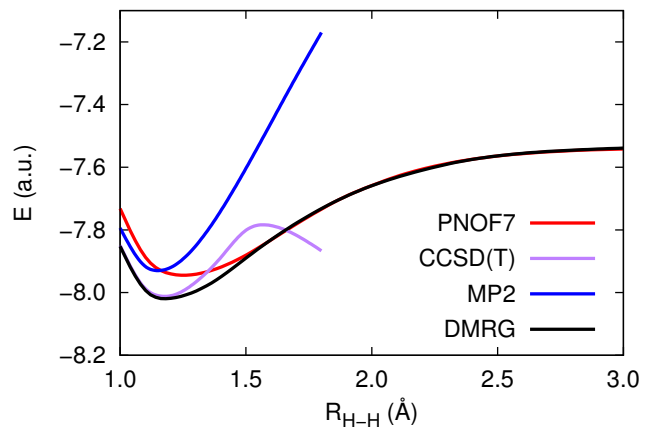}

\caption{\label{hydrogens-2d,6x6}Total energies for the symmetric dissociation
of the 4x4 hydrogen square lattice, in a.u., using PNOF7, CCSD(T),
MP2 and DMRG methods with the STO-6G basis set \cite{emls}.}
\end{figure}

In Fig. \ref{hydrogens-2d,6x6}, we show the energies obtained along
the symmetric dissociation of the 4x4 hydrogen square lattice by using
second-order Møller--Plesset perturbation theory (MP2), PNOF7, CCSD(T)
and DMRG methods with the STO-6G basis set \cite{emls}. The streching
of bonds has been done symmetrically in the two dimensions. CCSD(T) and MP2
calculations have been carried out by using the PSI4 electronic structure
package \cite{PSI4}, while the interface to CHEMPS2 by S. Wouters
et al. \cite{chemps2} has been employed for DMRG calculations. DMRG
is employed as reference for our calculations. In fact, despite
the limitations of tensor network state algorithms beyond 1D, the
latter are still very efficient to simulate short-range interactions
dominated systems \cite{dmrg1-PRB}, such as those encountered in
the minimal basis approach.

As expected, the single-reference methods MP2 and CCSD(T) fail in
multireference scenarios, where we observe large deviations of the
ONs from 0 and 1. CCSD(T) breaks down at $R_{H-H}>1.5\:\textrm{Å}$,
and does not even converge for long bond lengths $R_{H-H}>1.9\:\textrm{Å}$,
as previously observed in 1D hydrogen chains \cite{Hachman2006}.
For its part, MP2 provides reasonable energies slightly above CCSD(T)
around the equilibrium distance, but yields too high energies over
long distances, so it cannot describe the bond-breaking process either.

On the contrary, PNOF7 produces a qualitatively correct dissociation
curve. PNOF7 performs similar to MP2 around the equilibrium region, and it 
approaches reference DMRG results as the bond distance increases. It is worth 
noting the agreement at $R_{H-H}\geq1.5\:\textrm{Å}$, and specially at the 
dissociation limit. PNOF7 does not show convergence issues at any bond distance,
so it can be easily employed to describe bond-breaking processes in
presence of strong correlation effects, regardless of the dimensionality
of the system. If more accuracy is required around the equilibrium region, 
we must add the dynamic correlation to PNOF7. The recently proposed NOF-MP2 method 
\cite{Piris2017,Piris2018b} can be used to recover a significant part of the missing 
correlation. In fact, at the DMRG equilibrium distance of $R_{H-H}=1.18\:\textrm{Å}$, 
NOF-MP2 reduces the error by 0.04 a.u., which is half the difference between DMRG and 
PNOF7 at this distance.

\begin{table}[h]
\centering{}\caption{\label{table-IP-hydrogen}PNOF7 electron detachment energies, in a.u.,
($E^{S=1/2}-E^{S=0}$) for the chain (1D) and square lattice (2D)
composed of 16 hydrogen atoms with STO-6G basis set \cite{emls}.\medskip{}
}
\begin{tabular}{|c||c|c|c|c|c|c|c|}
\hline 
$R\left(\textrm{Å}\right)$ & 0.6 & 0.8 & 1.0 & 1.4 & 2.0 & 3.0 & 4.0\tabularnewline
\hline 
1D & 0.03 & 0.16 & 0.21 & 0.29 & 0.37 & 0.45 & 0.47\tabularnewline
\hline 
2D & 0.02 & 0.12 & 0.14 & 0.21 & 0.33 & 0.44 & 0.47\tabularnewline
\hline 
\end{tabular}
\end{table}

In Table \ref{table-IP-hydrogen}, we report the EDE along the dissociation
curve of the 4x4 hydrogen square lattice obtained as the energy difference
$E^{S=1/2}-E^{S=0}$. In contrast to the results obtained for the
Hubbard model, the EDE are positive along the whole dissociation curve
so can be interpreted as ionization energies. EDE take larger values
as non-dynamic correlation increases at large bond distances. These
results leave us the following conclusions: (1) long-range non-dynamic
correlation effects are crucial for the study of different spin multiplicities
and they must be carefully introduced in the Hubbard model by additional
terms in the Hamiltonian, and (2) EDE show the same trend regardless
the dimensionality of the system, so EDE for 1D or 2D system with
16 electrons are comparable as shown in Table \ref{table-IP-hydrogen}.

\section{Closing Remarks}

In the present paper, it was proved that PNOF7 is an efficient
method to describe strongly correlated electrons in two dimensions
with a mean-field computational scaling. Two models were extensively
investigated, namely the Hubbard model and the square lattice of hydrogens.
We studied the singlet ground-state, as well as the doublet mixed
quantum state that is obtained by extracting an electron from the
system.

It was shown that the performance of the present RDM functional approximation
is comparable to that of the state-of-the-art numerical methods such
as AFQMC or DMRG for the Hubbard model defined on 2D square lattices.
This agreement was confirmed for many filling situations and sizes
up to 144 electrons, from weak to strong correlation. Unlike other
RDM methods, PNOF7 showed no dependence on dimensionality, and we
obtained the same accuracy in two dimensions as that achieved in one
dimension. An outstanding feature is that the performance of PNOF7
does not deteriorate with the increasing size of the system. The reliability
of our energies was verified by calculations of the double occupancy
of sites that are known to be sensitive to the functional used in
the Hubbard model. No difficulties were observed in converging PNOF7
away from half-filling. It was corroborated that the localization
of the electrons that takes place at large $U/t$ values favors the
removal of an electron.

In the case of the hydrogen lattice, which resembles the 2D Hubbard
model with long-range interactions, PNOF7 showed good convergence
properties for any bond distance. The results obtained are close to
the DMRG reference values \LyXZeroWidthSpace \LyXZeroWidthSpace throughout
the symmetric dissociation curve and especially at the dissociation
limit. It was observed that the calculated electron detachment energies
(ionization energies) increase with the bond distance. It can be concluded
that the hydrogen lattice model can be employed to describe bond-breaking
processes in presence of strong correlation effects, regardless of
the dimensionality of the system.

In view of the accurate results obtained in this 2D study, together
with those previously obtained for strongly correlated electrons in
1D, we conclude that PNOF7 is a simple, cheap and efficient computational
method for the study of electron correlation in 1D and 2D systems.
The natural orbital functional theory is not only a promising
method of quantum chemistry, but also an emerging method for the condensed
matter physics.

\subsection*{Acknowledgments}

\textcolor{black}{Financial support comes from MCIU/AEI/FEDER, UE
(PGC2018-097529-B-100) and }Eusko Jaurlaritza (Ref. IT1254-19)\textcolor{black}{.
The authors thank for technical and human support provided by IZO-SGI
SGIker of UPV/EHU and European funding (ERDF and ESF). }I.M. is grateful
to Vice-Rectory for research of the UPV/EHU for the Ph. D. grant (PIF//15/043).


\begin{thebibliography}{41}
\expandafter\ifx\csname natexlab\endcsname\relax\def\natexlab#1{#1}\fi
\expandafter\ifx\csname bibnamefont\endcsname\relax
  \def\bibnamefont#1{#1}\fi
\expandafter\ifx\csname bibfnamefont\endcsname\relax
  \def\bibfnamefont#1{#1}\fi
\expandafter\ifx\csname citenamefont\endcsname\relax
  \def\citenamefont#1{#1}\fi
\expandafter\ifx\csname url\endcsname\relax
  \def\url#1{\texttt{#1}}\fi
\expandafter\ifx\csname urlprefix\endcsname\relax\def\urlprefix{URL }\fi
\providecommand{\bibinfo}[2]{#2}
\providecommand{\eprint}[2][]{\url{#2}}

\bibitem[{\citenamefont{Jingcheng et~al.}(2019)\citenamefont{Jingcheng, Sanz,
  Corso, Jang~Choi, Pe\~na, Frederiksen, and Ignacio~Pascual}}]{materials-2}
\bibinfo{author}{\bibfnamefont{L.}~\bibnamefont{Jingcheng}},
  \bibinfo{author}{\bibfnamefont{S.}~\bibnamefont{Sanz}},
  \bibinfo{author}{\bibfnamefont{M.}~\bibnamefont{Corso}},
  \bibinfo{author}{\bibfnamefont{D.}~\bibnamefont{Jang~Choi}},
  \bibinfo{author}{\bibfnamefont{D.}~\bibnamefont{Pe\~na}},
  \bibinfo{author}{\bibfnamefont{T.}~\bibnamefont{Frederiksen}},
  \bibnamefont{and}
  \bibinfo{author}{\bibfnamefont{J.}~\bibnamefont{Ignacio~Pascual}},
  \bibinfo{journal}{Nat. Commun} \textbf{\bibinfo{volume}{10}}
  (\bibinfo{year}{2019}).

\bibitem[{\citenamefont{Zheng et~al.}(2017)\citenamefont{Zheng, Chung, Corboz,
  Ehlers, Qin, Noack, Shi, White, Zhang, and Chan}}]{Zheng1155}
\bibinfo{author}{\bibfnamefont{B.-X.} \bibnamefont{Zheng}},
  \bibinfo{author}{\bibfnamefont{C.-M.} \bibnamefont{Chung}},
  \bibinfo{author}{\bibfnamefont{P.}~\bibnamefont{Corboz}},
  \bibinfo{author}{\bibfnamefont{G.}~\bibnamefont{Ehlers}},
  \bibinfo{author}{\bibfnamefont{M.-P.} \bibnamefont{Qin}},
  \bibinfo{author}{\bibfnamefont{R.~M.} \bibnamefont{Noack}},
  \bibinfo{author}{\bibfnamefont{H.}~\bibnamefont{Shi}},
  \bibinfo{author}{\bibfnamefont{S.~R.} \bibnamefont{White}},
  \bibinfo{author}{\bibfnamefont{S.}~\bibnamefont{Zhang}}, \bibnamefont{and}
  \bibinfo{author}{\bibfnamefont{G.~K.-L.} \bibnamefont{Chan}},
  \bibinfo{journal}{Science} \textbf{\bibinfo{volume}{358}},
  \bibinfo{pages}{1155} (\bibinfo{year}{2017}).

\bibitem[{\citenamefont{Qin et~al.}(2016)\citenamefont{Qin, Shi, and
  Zhang}}]{Mingpu}
\bibinfo{author}{\bibfnamefont{M.}~\bibnamefont{Qin}},
  \bibinfo{author}{\bibfnamefont{H.}~\bibnamefont{Shi}}, \bibnamefont{and}
  \bibinfo{author}{\bibfnamefont{S.}~\bibnamefont{Zhang}},
  \bibinfo{journal}{Phys. Rev. B} \textbf{\bibinfo{volume}{94}},
  \bibinfo{pages}{085103} (\bibinfo{year}{2016}).

\bibitem[{\citenamefont{Zheng and Chan}(2016)}]{dmet-prb}
\bibinfo{author}{\bibfnamefont{B.-X.} \bibnamefont{Zheng}} \bibnamefont{and}
  \bibinfo{author}{\bibfnamefont{G.~K.-L.} \bibnamefont{Chan}},
  \bibinfo{journal}{Phys. Rev. B} \textbf{\bibinfo{volume}{93}},
  \bibinfo{pages}{035126} (\bibinfo{year}{2016}).

\bibitem[{\citenamefont{Raghu et~al.}(2010)\citenamefont{Raghu, Kivelson, and
  Scalapino}}]{renormalization-group}
\bibinfo{author}{\bibfnamefont{S.}~\bibnamefont{Raghu}},
  \bibinfo{author}{\bibfnamefont{S.~A.} \bibnamefont{Kivelson}},
  \bibnamefont{and} \bibinfo{author}{\bibfnamefont{D.~J.}
  \bibnamefont{Scalapino}}, \bibinfo{journal}{Phys. Rev. B}
  \textbf{\bibinfo{volume}{81}}, \bibinfo{pages}{224505}
  (\bibinfo{year}{2010}).

\bibitem[{\citenamefont{Verstichel et~al.}(2014)\citenamefont{Verstichel,
  Poelmans, {De Baerdemacker}, Wouters, and {Van Neck}}}]{Verstichel2014}
\bibinfo{author}{\bibfnamefont{B.}~\bibnamefont{Verstichel}},
  \bibinfo{author}{\bibfnamefont{W.}~\bibnamefont{Poelmans}},
  \bibinfo{author}{\bibfnamefont{S.}~\bibnamefont{{De Baerdemacker}}},
  \bibinfo{author}{\bibfnamefont{S.}~\bibnamefont{Wouters}}, \bibnamefont{and}
  \bibinfo{author}{\bibfnamefont{D.}~\bibnamefont{{Van Neck}}},
  \bibinfo{journal}{Eur. Phys. J. B} \textbf{\bibinfo{volume}{87}},
  \bibinfo{pages}{59} (\bibinfo{year}{2014}).

\bibitem[{\citenamefont{Ehlers et~al.}(2015)\citenamefont{Ehlers, S\'olyom,
  Legeza, and Noack}}]{dmrg1-PRB}
\bibinfo{author}{\bibfnamefont{G.}~\bibnamefont{Ehlers}},
  \bibinfo{author}{\bibfnamefont{J.}~\bibnamefont{S\'olyom}},
  \bibinfo{author}{\bibfnamefont{O.}~\bibnamefont{Legeza}}, \bibnamefont{and}
  \bibinfo{author}{\bibfnamefont{R.~M.} \bibnamefont{Noack}},
  \bibinfo{journal}{Phys. Rev. B} \textbf{\bibinfo{volume}{92}},
  \bibinfo{pages}{235116} (\bibinfo{year}{2015}).

\bibitem[{\citenamefont{and LeBlanc et~al.}(2015)\citenamefont{and LeBlanc,
  Antipov, Becca, Bulik, Chan, Chung, Deng, Ferrero, Henderson,
  Jim{\'{e}}nez-Hoyos et~al.}}]{PHYSREVX_2Dhubbard}
\bibinfo{author}{\bibfnamefont{J.~P.~F.} \bibnamefont{and LeBlanc}},
  \bibinfo{author}{\bibfnamefont{A.~E.} \bibnamefont{Antipov}},
  \bibinfo{author}{\bibfnamefont{F.}~\bibnamefont{Becca}},
  \bibinfo{author}{\bibfnamefont{I.~W.} \bibnamefont{Bulik}},
  \bibinfo{author}{\bibfnamefont{G.~K.-L.} \bibnamefont{Chan}},
  \bibinfo{author}{\bibfnamefont{C.-M.} \bibnamefont{Chung}},
  \bibinfo{author}{\bibfnamefont{Y.}~\bibnamefont{Deng}},
  \bibinfo{author}{\bibfnamefont{M.}~\bibnamefont{Ferrero}},
  \bibinfo{author}{\bibfnamefont{T.~M.} \bibnamefont{Henderson}},
  \bibinfo{author}{\bibfnamefont{C.~A.} \bibnamefont{Jim{\'{e}}nez-Hoyos}},
  \bibnamefont{et~al.}, \bibinfo{journal}{Phys. Rev. X}
  \textbf{\bibinfo{volume}{5}}, \bibinfo{pages}{1} (\bibinfo{year}{2015}).

\bibitem[{\citenamefont{Pernal}(2015)}]{Pernal_2015_new_journal}
\bibinfo{author}{\bibfnamefont{K.}~\bibnamefont{Pernal}}, \bibinfo{journal}{New
  J. Phys.} \textbf{\bibinfo{volume}{17}}, \bibinfo{pages}{111001}
  (\bibinfo{year}{2015}).

\bibitem[{\citenamefont{Anderson et~al.}(2013)\citenamefont{Anderson, Nakata,
  Igarashi, Fujisawa, and Yamashita}}]{ANDERSON201322}
\bibinfo{author}{\bibfnamefont{J.~S.} \bibnamefont{Anderson}},
  \bibinfo{author}{\bibfnamefont{M.}~\bibnamefont{Nakata}},
  \bibinfo{author}{\bibfnamefont{R.}~\bibnamefont{Igarashi}},
  \bibinfo{author}{\bibfnamefont{K.}~\bibnamefont{Fujisawa}}, \bibnamefont{and}
  \bibinfo{author}{\bibfnamefont{M.}~\bibnamefont{Yamashita}},
  \bibinfo{journal}{Comp. Theor. Chem.} \textbf{\bibinfo{volume}{1003}},
  \bibinfo{pages}{22 } (\bibinfo{year}{2013}).

\bibitem[{\citenamefont{Pernal and Giesbertz}(2016)}]{Pernal2016}
\bibinfo{author}{\bibfnamefont{K.}~\bibnamefont{Pernal}} \bibnamefont{and}
  \bibinfo{author}{\bibfnamefont{K.~J.~H.} \bibnamefont{Giesbertz}},
  \bibinfo{journal}{Top Curr Chem} \textbf{\bibinfo{volume}{368}},
  \bibinfo{pages}{125} (\bibinfo{year}{2016}).

\bibitem[{\citenamefont{Johnson and Becke}(2017)}]{becke_dft_strong}
\bibinfo{author}{\bibfnamefont{E.~R.} \bibnamefont{Johnson}} \bibnamefont{and}
  \bibinfo{author}{\bibfnamefont{A.~D.} \bibnamefont{Becke}},
  \bibinfo{journal}{J. Chem. Phys.} \textbf{\bibinfo{volume}{146}},
  \bibinfo{pages}{211105} (\bibinfo{year}{2017}).

\bibitem[{\citenamefont{Su et~al.}(2018)\citenamefont{Su, Li, and
  Yang}}]{Su_DFT_strong}
\bibinfo{author}{\bibfnamefont{N.~Q.} \bibnamefont{Su}},
  \bibinfo{author}{\bibfnamefont{C.}~\bibnamefont{Li}}, \bibnamefont{and}
  \bibinfo{author}{\bibfnamefont{W.}~\bibnamefont{Yang}},
  \bibinfo{journal}{Proc. Natl. Acad. Sci.} \textbf{\bibinfo{volume}{115}},
  \bibinfo{pages}{9678} (\bibinfo{year}{2018}).

\bibitem[{\citenamefont{Carrascal et~al.}(2015)\citenamefont{Carrascal, Ferrer,
  Smith, and Burke}}]{Carrascal}
\bibinfo{author}{\bibfnamefont{D.~J.} \bibnamefont{Carrascal}},
  \bibinfo{author}{\bibfnamefont{J.}~\bibnamefont{Ferrer}},
  \bibinfo{author}{\bibfnamefont{J.~C.} \bibnamefont{Smith}}, \bibnamefont{and}
  \bibinfo{author}{\bibfnamefont{K.}~\bibnamefont{Burke}}, \bibinfo{journal}{J.
  Phys.: Condens. Matter} \textbf{\bibinfo{volume}{27}},
  \bibinfo{pages}{393001} (\bibinfo{year}{2015}).

\bibitem[{\citenamefont{Shinohara et~al.}(2015)\citenamefont{Shinohara, Sharma,
  Dewhurst, Shallcross, Lathiotakis, and Gross}}]{Shinohara_2015}
\bibinfo{author}{\bibfnamefont{Y.}~\bibnamefont{Shinohara}},
  \bibinfo{author}{\bibfnamefont{S.}~\bibnamefont{Sharma}},
  \bibinfo{author}{\bibfnamefont{J.~K.} \bibnamefont{Dewhurst}},
  \bibinfo{author}{\bibfnamefont{S.}~\bibnamefont{Shallcross}},
  \bibinfo{author}{\bibfnamefont{N.~N.} \bibnamefont{Lathiotakis}},
  \bibnamefont{and} \bibinfo{author}{\bibfnamefont{E.~K.~U.}
  \bibnamefont{Gross}}, \bibinfo{journal}{New J. Phys.}
  \textbf{\bibinfo{volume}{17}}, \bibinfo{pages}{093038}
  (\bibinfo{year}{2015}).

\bibitem[{\citenamefont{Mitxelena et~al.}(2017)\citenamefont{Mitxelena, Piris,
  and Mayorga}}]{Mitxelena2017a}
\bibinfo{author}{\bibfnamefont{I.}~\bibnamefont{Mitxelena}},
  \bibinfo{author}{\bibfnamefont{M.}~\bibnamefont{Piris}}, \bibnamefont{and}
  \bibinfo{author}{\bibfnamefont{M.} \bibnamefont{Rodr\'{i}guez-Mayorga}},
  \bibinfo{journal}{J. Phys. Condens. Matter} \textbf{\bibinfo{volume}{29}},
  \bibinfo{pages}{425602} (\bibinfo{year}{2017}).

\bibitem[{\citenamefont{Mitxelena
  et~al.}(2018{\natexlab{a}})\citenamefont{Mitxelena, Piris, and
  Rodr\'{i}guez-Mayorga}}]{Mitxelena2018b}
\bibinfo{author}{\bibfnamefont{I.}~\bibnamefont{Mitxelena}},
  \bibinfo{author}{\bibfnamefont{M.}~\bibnamefont{Piris}}, \bibnamefont{and}
  \bibinfo{author}{\bibfnamefont{M.}~\bibnamefont{Rodr\'{i}guez-Mayorga}},
  \bibinfo{journal}{J. Phys. Condens. Matter} \textbf{\bibinfo{volume}{30}},
  \bibinfo{pages}{089501} (\bibinfo{year}{2018}{\natexlab{a}}).

\bibitem[{\citenamefont{Mitxelena
  et~al.}(2018{\natexlab{b}})\citenamefont{Mitxelena, Rodr{\'{i}}guez-Mayorga,
  and Piris}}]{mitxelena2018a}
\bibinfo{author}{\bibfnamefont{I.}~\bibnamefont{Mitxelena}},
  \bibinfo{author}{\bibfnamefont{M.}~\bibnamefont{Rodr{\'{i}}guez-Mayorga}},
  \bibnamefont{and} \bibinfo{author}{\bibfnamefont{M.}~\bibnamefont{Piris}},
  \bibinfo{journal}{Eur. Phys. J. B} \textbf{\bibinfo{volume}{91}},
  \bibinfo{pages}{109} (\bibinfo{year}{2018}{\natexlab{b}}).

\bibitem[{\citenamefont{Schilling and Schilling}(2019)}]{Schilling2019}
\bibinfo{author}{\bibfnamefont{C.}~\bibnamefont{Schilling}} \bibnamefont{and}
  \bibinfo{author}{\bibfnamefont{R.}~\bibnamefont{Schilling}},
  \bibinfo{journal}{Phys. Rev. Lett.} \textbf{\bibinfo{volume}{122}},
  \bibinfo{pages}{013001} (\bibinfo{year}{2019}).
  
\bibitem[{\citenamefont{Mitxelena et~al.}(2019)\citenamefont{Mitxelena, Piris,
  and Ugalde}}]{MITXELENA2019}
\bibinfo{author}{\bibfnamefont{I.}~\bibnamefont{Mitxelena}},
  \bibinfo{author}{\bibfnamefont{M.}~\bibnamefont{Piris}}, \bibnamefont{and}
  \bibinfo{author}{\bibfnamefont{J.~M.} \bibnamefont{Ugalde}}, in
  \emph{\bibinfo{booktitle}{State of The Art of Molecular Electronic Structure
  Computations: Correlation Methods, Basis Sets and More}}, edited by
  \bibinfo{editor}{\bibfnamefont{L.~U.} \bibnamefont{Ancarani}}
  \bibnamefont{and} \bibinfo{editor}{\bibfnamefont{P.~E.} \bibnamefont{Hoggan}}
  (\bibinfo{publisher}{Academic Press}, \bibinfo{year}{2019}),
  vol.~\bibinfo{volume}{79} of \emph{\bibinfo{series}{Advances in Quantum
  Chemistry}}, pp. \bibinfo{pages}{155 -- 177}.

\bibitem[{\citenamefont{Sauban\`ere and Pastor}(2011)}]{Saubanere-prb-2011}
\bibinfo{author}{\bibfnamefont{M.}~\bibnamefont{Sauban\`ere}} \bibnamefont{and}
  \bibinfo{author}{\bibfnamefont{G.~M.} \bibnamefont{Pastor}},
  \bibinfo{journal}{Phys. Rev. B} \textbf{\bibinfo{volume}{84}},
  \bibinfo{pages}{035111} (\bibinfo{year}{2011}).

\bibitem[{\citenamefont{Piris}(2007)}]{Piris2007}
\bibinfo{author}{\bibfnamefont{M.}~\bibnamefont{Piris}}, in
  \emph{\bibinfo{booktitle}{Reduced-Density-Matrix Mechanics: with applications
  to many-electron atoms and molecules}}, edited by
  \bibinfo{editor}{\bibfnamefont{D.~A.} \bibnamefont{Mazziotti}}
  (\bibinfo{publisher}{John Wiley and Sons}, \bibinfo{address}{Hoboken, New
  Jersey, USA}, \bibinfo{year}{2007}), chap.~\bibinfo{chapter}{14}, pp.
  \bibinfo{pages}{387--427}.

\bibitem[{\citenamefont{Piris and Ugalde}(2014)}]{Piris2014a}
\bibinfo{author}{\bibfnamefont{M.}~\bibnamefont{Piris}} \bibnamefont{and}
  \bibinfo{author}{\bibfnamefont{J.~M.} \bibnamefont{Ugalde}},
  \bibinfo{journal}{Int. J. Quantum Chem.} \textbf{\bibinfo{volume}{114}},
  \bibinfo{pages}{1169} (\bibinfo{year}{2014}).

\bibitem[{\citenamefont{Piris}(2017)}]{Piris2017}
\bibinfo{author}{\bibfnamefont{M.}~\bibnamefont{Piris}},
  \bibinfo{journal}{Phys. Rev. Lett.} \textbf{\bibinfo{volume}{119}},
  \bibinfo{pages}{063002} (\bibinfo{year}{2017}).

\bibitem[{\citenamefont{Motta et~al.}(2017)\citenamefont{Motta, Ceperley, Chan,
  Gomez, Gull, Guo, Jim\'enez-Hoyos, Lan, Li, Ma et~al.}}]{review_Hchain}
\bibinfo{author}{\bibfnamefont{M.}~\bibnamefont{Motta}},
  \bibinfo{author}{\bibfnamefont{D.~M.} \bibnamefont{Ceperley}},
  \bibinfo{author}{\bibfnamefont{G.~K.-L.} \bibnamefont{Chan}},
  \bibinfo{author}{\bibfnamefont{J.~A.} \bibnamefont{Gomez}},
  \bibinfo{author}{\bibfnamefont{E.}~\bibnamefont{Gull}},
  \bibinfo{author}{\bibfnamefont{S.}~\bibnamefont{Guo}},
  \bibinfo{author}{\bibfnamefont{C.~A.} \bibnamefont{Jim\'enez-Hoyos}},
  \bibinfo{author}{\bibfnamefont{T.~N.} \bibnamefont{Lan}},
  \bibinfo{author}{\bibfnamefont{J.}~\bibnamefont{Li}},
  \bibinfo{author}{\bibfnamefont{F.}~\bibnamefont{Ma}}, \bibnamefont{et~al.},
  \bibinfo{journal}{Phys. Rev. X} \textbf{\bibinfo{volume}{7}},
  \bibinfo{pages}{031059} (\bibinfo{year}{2017}).

\bibitem[{\citenamefont{Coleman}(1963)}]{Coleman1963}
\bibinfo{author}{\bibfnamefont{A.~J.} \bibnamefont{Coleman}},
  \bibinfo{journal}{Rev. Mod. Phys.} \textbf{\bibinfo{volume}{35}},
  \bibinfo{pages}{668} (\bibinfo{year}{1963}).

\bibitem[{\citenamefont{{M. Piris}}(2019)}]{Piris2019}
\bibinfo{author}{\bibnamefont{{M. Piris}}}, \bibinfo{journal}{Phys. Rev. A}
  \textbf{\bibinfo{volume}{100}}, \bibinfo{pages}{032508}
  (\bibinfo{year}{2019}).

\bibitem[{\citenamefont{Mazziotti}(2012)}]{Mazziotti2012a}
\bibinfo{author}{\bibfnamefont{D.~A.} \bibnamefont{Mazziotti}},
  \bibinfo{journal}{Phys. Rev. Lett.} \textbf{\bibinfo{volume}{108}},
  \bibinfo{pages}{263002} (\bibinfo{year}{2012}).

\bibitem[{\citenamefont{Piris}(2018{\natexlab{a}})}]{Piris2018}
\bibinfo{author}{\bibfnamefont{M.}~\bibnamefont{Piris}}, in
  \emph{\bibinfo{booktitle}{Many-body approaches at different scales: a tribute
  to N. H. March on the occasion of his 90th birthday}}, edited by
  \bibinfo{editor}{\bibfnamefont{G.~G.~N.} \bibnamefont{Angilella}}
  \bibnamefont{and} \bibinfo{editor}{\bibfnamefont{C.}~\bibnamefont{Amovilli}}
  (\bibinfo{publisher}{Springer}, \bibinfo{address}{New York},
  \bibinfo{year}{2018}{\natexlab{a}}), chap.~\bibinfo{chapter}{22}, pp.
  \bibinfo{pages}{283--300}.

\bibitem[{\citenamefont{Piris}(2018{\natexlab{b}})}]{Piris2018a}
\bibinfo{author}{\bibfnamefont{M.}~\bibnamefont{Piris}}, in
  \emph{\bibinfo{booktitle}{Theoretical and Quantum Chemistry at the Dawn of
  the 21st Century}}, edited by
  \bibinfo{editor}{\bibfnamefont{T.}~\bibnamefont{Chakraborty}}
  \bibnamefont{and}
  \bibinfo{editor}{\bibfnamefont{R.}~\bibnamefont{Carb{\'{o}}-Dorca}}
  (\bibinfo{publisher}{Apple Academic Press}, \bibinfo{address}{New Jersey},
  \bibinfo{year}{2018}{\natexlab{b}}), chap.~\bibinfo{chapter}{22}, pp.
  \bibinfo{pages}{593--620}.

\bibitem[{\citenamefont{Piris and Ugalde}(2009)}]{Piris2009a}
\bibinfo{author}{\bibfnamefont{M.}~\bibnamefont{Piris}} \bibnamefont{and}
  \bibinfo{author}{\bibfnamefont{J.~M.} \bibnamefont{Ugalde}},
  \bibinfo{journal}{J. Comput. Chem.} \textbf{\bibinfo{volume}{30}},
  \bibinfo{pages}{2078} (\bibinfo{year}{2009}).

\bibitem[{\citenamefont{Verstichel et~al.}(2012)\citenamefont{Verstichel, van
  Aggelen, Poelmans, and Van~Neck}}]{verstichel-prl}
\bibinfo{author}{\bibfnamefont{B.}~\bibnamefont{Verstichel}},
  \bibinfo{author}{\bibfnamefont{H.}~\bibnamefont{van Aggelen}},
  \bibinfo{author}{\bibfnamefont{W.}~\bibnamefont{Poelmans}}, \bibnamefont{and}
  \bibinfo{author}{\bibfnamefont{D.}~\bibnamefont{Van~Neck}},
  \bibinfo{journal}{Phys. Rev. Lett.} \textbf{\bibinfo{volume}{108}},
  \bibinfo{pages}{213001} (\bibinfo{year}{2012}).

\bibitem[{\citenamefont{Rubin and Mazziotti}(2015)}]{rubin_mazziotti_2015}
\bibinfo{author}{\bibfnamefont{N.~C.} \bibnamefont{Rubin}} \bibnamefont{and}
  \bibinfo{author}{\bibfnamefont{D.~A.} \bibnamefont{Mazziotti}},
  \bibinfo{journal}{J. Phys. Chem. C} \textbf{\bibinfo{volume}{119}},
  \bibinfo{pages}{14706} (\bibinfo{year}{2015}).

\bibitem[{\citenamefont{Rubin and Mazziotti}(2014)}]{Rubin_mazziotti_2014}
\bibinfo{author}{\bibfnamefont{N.~C.} \bibnamefont{Rubin}} \bibnamefont{and}
  \bibinfo{author}{\bibfnamefont{D.~A.} \bibnamefont{Mazziotti}},
  \bibinfo{journal}{Theor. Chem. Acc.} \textbf{\bibinfo{volume}{133}},
  \bibinfo{pages}{1492} (\bibinfo{year}{2014}).

\bibitem[{\citenamefont{Gritsenko and Pernal}(2019)}]{Gritsenko_pra2019}
\bibinfo{author}{\bibfnamefont{O.~V.} \bibnamefont{Gritsenko}}
  \bibnamefont{and} \bibinfo{author}{\bibfnamefont{K.}~\bibnamefont{Pernal}},
  \bibinfo{journal}{Phys. Rev. A} \textbf{\bibinfo{volume}{100}},
  \bibinfo{pages}{012509} (\bibinfo{year}{2019}).

\bibitem[{\citenamefont{Mitxelena and Piris}(2020 (in press))}]{mitxelena2019c}
\bibinfo{author}{\bibfnamefont{I.}~\bibnamefont{Mitxelena}} \bibnamefont{and}
  \bibinfo{author}{\bibfnamefont{M.}~\bibnamefont{Piris}},
  \bibinfo{journal}{J. Phys: Condens. Matter DOI: 10.1088/1361-648X/ab6d11}
  (\bibinfo{year}{2020}).

\bibitem[{\citenamefont{P.~Pritchard et~al.}(2019 (in
  preparation))\citenamefont{P.~Pritchard, Altarawy, Didier, Gibson, and
  Windus}}]{emls}
\bibinfo{author}{\bibfnamefont{B.}~\bibnamefont{P.~Pritchard}},
  \bibinfo{author}{\bibfnamefont{D.}~\bibnamefont{Altarawy}},
  \bibinfo{author}{\bibfnamefont{B.}~\bibnamefont{Didier}},
  \bibinfo{author}{\bibfnamefont{T.~D.} \bibnamefont{Gibson}},
  \bibnamefont{and} \bibinfo{author}{\bibfnamefont{T.~L.} \bibnamefont{Windus}}
  (\bibinfo{year}{2019 (in preparation)}).

\bibitem[{\citenamefont{Parrish et~al.}(2017)\citenamefont{Parrish, Burns,
  Smith, Simmonett, DePrince, Hohenstein, Bozkaya, Sokolov, Di~Remigio, Richard
  et~al.}}]{PSI4}
\bibinfo{author}{\bibfnamefont{R.~M.} \bibnamefont{Parrish}},
  \bibinfo{author}{\bibfnamefont{L.~A.} \bibnamefont{Burns}},
  \bibinfo{author}{\bibfnamefont{D.~G.~A.} \bibnamefont{Smith}},
  \bibinfo{author}{\bibfnamefont{A.~C.} \bibnamefont{Simmonett}},
  \bibinfo{author}{\bibfnamefont{A.~E.} \bibnamefont{DePrince}},
  \bibinfo{author}{\bibfnamefont{E.~G.} \bibnamefont{Hohenstein}},
  \bibinfo{author}{\bibfnamefont{U.}~\bibnamefont{Bozkaya}},
  \bibinfo{author}{\bibfnamefont{A.~Y.} \bibnamefont{Sokolov}},
  \bibinfo{author}{\bibfnamefont{R.}~\bibnamefont{Di~Remigio}},
  \bibinfo{author}{\bibfnamefont{R.~M.} \bibnamefont{Richard}},
  \bibnamefont{et~al.}, \bibinfo{journal}{J. Chem. Theory Comput.}
  \textbf{\bibinfo{volume}{13}}, \bibinfo{pages}{3185} (\bibinfo{year}{2017}).

\bibitem[{\citenamefont{Wouters et~al.}(2014)\citenamefont{Wouters, Poelmans,
  Ayers, and Neck}}]{chemps2}
\bibinfo{author}{\bibfnamefont{S.}~\bibnamefont{Wouters}},
  \bibinfo{author}{\bibfnamefont{W.}~\bibnamefont{Poelmans}},
  \bibinfo{author}{\bibfnamefont{P.~W.} \bibnamefont{Ayers}}, \bibnamefont{and}
  \bibinfo{author}{\bibfnamefont{D.~V.} \bibnamefont{Neck}},
  \bibinfo{journal}{Comput. Phys. Commun.} \textbf{\bibinfo{volume}{185}},
  \bibinfo{pages}{1501 } (\bibinfo{year}{2014}).

\bibitem[{\citenamefont{Hachman et~al.}(2006)\citenamefont{Hachman, Cardoen,
  and Chan}}]{Hachman2006}
\bibinfo{author}{\bibfnamefont{J.}~\bibnamefont{Hachman}},
  \bibinfo{author}{\bibfnamefont{W.}~\bibnamefont{Cardoen}}, \bibnamefont{and}
  \bibinfo{author}{\bibfnamefont{G.~K.-L.} \bibnamefont{Chan}},
  \bibinfo{journal}{J. Chem. Phys.} \textbf{\bibinfo{volume}{125}}
  (\bibinfo{year}{2006}).

\bibitem[{\citenamefont{{M. Piris}}(2018)}]{Piris2018b}
\bibinfo{author}{\bibnamefont{{M. Piris}}}, \bibinfo{journal}{Phys. Rev. A}
  \textbf{\bibinfo{volume}{98}}, \bibinfo{pages}{022504}
  (\bibinfo{year}{2018}).

\end{thebibliography}
\end{document}